\documentclass[10pt]{article}
\usepackage{epsf}
\usepackage[a4paper,lmargin=1cm]{geometry}
\usepackage{color}

\title{Non-minimal Quintessence: Dynamics and coincidence problem}
\author{Fatimah Shojai and Ali Shojai\\ Department of Physics, University of Tehran, Tehran, Iran.}
\date{}
\begin{document}
\maketitle
\begin{abstract}
Brans--Dicke scalar--tensor theory provides a conformally coupling of the scalar field with gravity in Einstein's frame. This model is equivalent to an interacting quintessence in which dark matter is coupled to dark energy. This provides a natural mechanism to alleviate the coincidence problem. We investigate the dynamics of this model and show that it leads to comparable dark energy and dark matter densities today.
\end{abstract}
\section{Introduction}
According to Einstein's theory of relativity, the observed late time acceleration\cite{Riess} of the universe can be successfully explained by introducing the dark energy  component\cite{sami} on the right hand side of the Einstein's equations. There are two main scenarios used to explain the nature of dark energy: a cosmological constant $\Lambda$\cite{carroll} or a scalar field, usually called quintessence\cite{quintessence}. The cosmological model that consists of a mixture of cosmological constant and cold dark matter is called $\Lambda$CDM and the other possibility leads to QCDM cosmological model. Although  $\Lambda$CDM model opens the possibility to fit it with the observational data, two questions arise: first, why the value of the cosmological constant is so small, and second, why it's energy density  begins to dominate now. This is known as coincidence problem\cite{coincidence}. 

Quintessential models of dark energy also provide an adequate fit to the cosmological data. {Moreover the existance of stable tracking solutions of the quintessence field shows the importance of this model in the investigation of the fine tuning problem. But it can not explain the present ratio of energy densities in the universe, i.e. the coincidence problem \cite{amen}. One possible way to make a model to overcome the coincidence and fine tuning problems, is to introduce a proper coupling term between the dark energy and dark matter.} The form of the interaction term is motivated from the phenomenological consideration or by requiring that the ratio of dark energy to dark matter densities, $r={\rho_\varphi}/{\rho_c}$, be stable under perturbations\cite{amen,2olivares}.

Introducing the interaction in QCDM model, can lead to a finite $r$ at present or causes this ratio varies slowly now, which is called soft coincidence\cite{campo}. Moreover it is required to have a smooth transition from matter dominated universe to the accelerated expanding universe. Denoting the coupling term by $Q$ in the conservation equations of the quintessence and cosmic dark fluid, one has:
\begin{equation}\label{rhocdot}
\dot{\rho}_{c}+3H\rho_{c}=Q
\end{equation}
\begin{equation}\label{rhophidot}
\dot{\rho}_{\varphi}+3H(1+\omega)\rho_{\varphi}=-Q
\end{equation}
in which $\rho_{c}$, $\rho_{\varphi}=\frac{1}{2}\dot{\varphi}^{2}+V(\varphi)$ and
$p_{\varphi}=\frac{1}{2}\dot{\varphi}^{2}-V(\varphi)$ are the energy density of cosmic dark fluid, the energy density and pressure density of the scalar field respectively. $\omega=p_{\varphi}/\rho_{\varphi}$ is the equation of state parameter for the scalar field and a dot over any quantity denotes derivation with respect to the cosmic time. 

A variety of interaction terms have been suggested in the literature\cite{amen,2olivares,campo,oliv,gab,Q,amend,guo,epl-khodemoon,amendo,bohmer,s1,3bohmer,6bohmer,energy}. One of the most important ones is  $Q=\lambda_1H\rho_c$  \cite{campo,gab,Q,amend}. Inserting it into the above equations, together with the Friedmann equations, yields  a monotonic decreasing function of the scale factor for $r$ which changes slowly at the present time, i.e. $\left .\dot{r}/r \right |_{t_0}\ll{H_0}$ \cite{gab}.  Furthermore assuming an accelerated universe and negative $\dot{r}$, together with satisfying the soft coincidence criterion, the functions $H(a)$, $V(\varphi)$ and the allowed range of $\lambda_1$ are determined. More general interaction term $Q=\lambda_1H\rho_c+\lambda_2H\rho_{\varphi}$ is investigated in \cite{oliv, gab}. This yields to the evolution from a stable value $r_+$ at early times to an unstable value $r_-$ at late times \cite{gab}. Also the constants $\lambda_1$ and $\lambda_2$ are constrained by the evolution of the background energy density.

It must be noted that the observational data are unable to discriminate between this interaction and the previous one. Also these two models are investigated in a more general manner in which the transfer rate of energy densities is determined by the properties of dark sector and in general is different from the expansion rate of the universe\cite{guo}.

In \cite{amend}, it is assumed that the scaling solutions (having a constant $r$ at all times) exist for the generic form of a non-minimally scalar-field Lagrangian. This leads to the coupling $Q=\beta\rho_{c}\dot{\varphi}$ {in the Einstein's frame}. It can be shown {\cite{amendo} that the proper sequence of cosmological epochs in the form of radiation-dominated, $\varphi$-matter-dominated and accelerated epoch is followed in this model.} Also the cosmological consequences of this kind of coupling is studied in \cite{epl-khodemoon,amendo,bohmer,s1} assuming an exponential potential for the quintessence field. 

The simplest model which can provide this particular interaction is { a non minimally coupled gravity in which the scalar field couples minimally to gravity but couples conformally to the matter fields \cite{amendo,3bohmer}. This can be obtained from the following action \cite{11boom}:}
\begin{equation}
S_{E}=\int d^{4}x
\sqrt{-g}[R-\frac{1}{2}(\nabla\varphi)^{2}
-V(\varphi)+\mathcal{L}_{c}(e^{-2\beta\varphi}g_{\mu\nu})+\mathcal{L}_{b}(g_{\mu\nu})].
\label{ac}
\end{equation}
{It is obvious that this model is conformally related to a non minimal gravity in the form of Brans-Dicke type lagrangian. The cosmological properties of this model is investigated in the literature. In \cite{tor} an observational analysis of non-minimally coupled theory and it's differences from Genaral relativity is presented. Also the authors of \cite{mata} discuss the attractive aspect of non minimal quintessence and show that after tracking era, the quintessence dominated epoch is started. The particle production of this model in inflationary epoch is studied by Sahni et. al. \cite{sahni}. Their analysis  leads to a spatially flat universe consisting of dark energy (in the form of
quanta of the non-minimally coupled scalar field created during
inflation) and dark matter. In this way it presents a new way of generating dark energy quantum mechanically\cite{chom}.

For the latter coupling, more general form of the potential and allowing $\beta$ to be a function $\beta(\varphi)$ can be found in \cite{6bohmer}}.

{In summury, the coupling of dark energy to dark matter shows that the dark energy can affect the expansion history of the universe and the change of the growth rate of matter perturbations in the structure formation stage. Therefore this makes tight bounds on the strength of coupling terms \cite{energy}. Also in \cite{Q}, the different forms of coupling and their aspects have been investigated and compared with the observational data. the result indicates that the parameter space, which includes the arbitrary coupling constants and the equation of state parameter, is greatly limited.}   

{This work rests on the latter coupling pointed above. Moreover} we use new variables, proposed in \cite{gab}, to simplify the analysis of the critical points of the non-minimal scalar field model.   

The outline of this paper is as follows. In the next section we shall study the critical points of the autonomous system of equations associated with the non--minimally coupled quintessence model. Then in the third section we disscus the dependence of dark energy--dark matter ratio on the redshift. Finally in the last section we shall give some conclusions.

\section{The model and its critical points}
Here we consider the quintessence field decays into the pressureless CDM according to the latter kind of coupling term discussed in the introduction, i.e. $Q=\beta\rho_c\dot{\varphi}$. We also include baryons' density as an independent conserved component such that:
\begin{equation}\label{rhobdot}
\dot{\rho}_{b}+3H\rho_{b}=0
\end{equation}
The conservation equations (\ref{rhocdot}), (\ref{rhophidot}) and (\ref{rhobdot}) can be obtained from the  action (\ref{ac}). 
It has to be noted that similar models are disscussed in \cite{amendo} (in which baryons plus CDM couples to dark energy component) and in \cite{bohmer} (neglecting baryons). Both these works assume an exponential potential for dark energy. Here we shall use a more direct way to investigate the model by { assuming that the equation of state parameter, $\omega$, is constant and then absorb it} in the governing equations via introducing new variables, and we have not consider any assumption about the special form of the potential.

Now, consider a spatially flat FRW universe occupied by the above three components. The cosmological equations of motion are (we have chosen the unit in which $8\pi G=c=1$):
\begin{equation}\label{friedmann}
H^{2}=\frac{1}{3}(\rho_{\varphi}+\rho_{b}+\rho_{c})
\end{equation}
\begin{equation}\label{friedmann1}
\dot{H}=-\frac{1}{2}\left[\rho_{\varphi}(1+\omega)+\rho_{b}+\rho_{c}\right]
\end{equation}
Defining the dimensionless variables:
\begin{equation}
x = \frac{\rho_{\varphi}}{3H^2}   \quad        y = \frac{ \rho_c}{3H^2} \quad       u = \frac{Q}{3 H^3} 
\label{eq:variables}
\end{equation}
and following the same method as  \cite{gab}, the conservation relations of dark sector, equations (\ref{rhocdot}) and (\ref{rhophidot}) can be written as:
\begin{eqnarray}
 x^\prime &=& 3 x \left( 1 - x \right) - \tilde{u} \,
 , \label{sysa} \\
 y^\prime &=& -3 xy + \tilde{u} \, 
\label{sysb}
\end{eqnarray}
in which $\tilde{u}=-{u}/{\omega}$ and a prime denotes derivation with respect to $N=-\omega\ln a$ (we assume the current value of the scale factor is one, $a_0=1$). Using the relation $\dot{\varphi}^2=(1+\omega)\rho_\varphi$ , one obtains:
\begin{equation}
\tilde{u}=\pm\tilde{\beta}y\sqrt{3(1+\omega)x}
\label{z} 
\end{equation}
where $\tilde{\beta}=\beta/|\omega|$ and the positive (negative) sign indicates that $\dot{\varphi}$ and $\omega$ have the opposite (same) signs. Here we consider $\tilde{\beta}$ to be a positive constant.
Inserting (\ref{z}) in (\ref{sysa}) and (\ref{sysb}), the critical points of the above dynamical system together their stability along the dynamical state of the universe are analyzed by means of eigenvalues method and the results are  summarized in table (\ref{tab}). 

\begin{table}
\begin{center}
\begin{tabular}{ccccccc}
\hline\hline Point & $(x,y)$ & Existence & Stable? & Acceleration? & $\Omega_b$ & $Q$ \\ 
\hline A & $(0,y_0)$ & all $\omega$, $\beta$ & no & no & $\neq 0$ & $0$ \\ 
\hline B Type I & $(1,0)$ & all $\omega$, $\beta$ & $\beta<\sqrt{\frac{3}{1+\omega}}|\omega|$ & $\omega<-\frac{1}{3}$ & $0$ & $0$ \\ 
\hline B Type II & $(1,0)$ & all $\omega$, $\beta$ & all $\beta$ & $\omega<-\frac{1}{3}$ & $0$ & $0$ \\ 
\hline C Type I & $(\frac{(1+\omega)\tilde{\beta}^2}{3},1-\frac{(1+\omega)\tilde{\beta}^2}{3})$ & all $\omega$, $\beta$ & $\beta>\sqrt{\frac{3}{1+\omega}}|\omega|$ & $|\omega|<\frac{\beta^2}{1+\beta^2}$ & $0$ & $-3H^3\omega\tilde{\beta}^2(1+\omega)$ \\ 
\hline \hline
\end{tabular}
\caption{Critical points of the model and their properties.}
\label{tab} 
\end{center}
\end{table}

The critical points are labeled by $A$, $B$ and $C$ and the type $I$ ($II$) solution corresponds to the opposite (same) sign of $\dot\varphi$ and $\omega$. The first critical point, $A$, is $(x=0,y=y_0)$ which results in $\tilde{u}=0$ and thus there is neither scalar field nor interaction. This point represents the baryon-cold dark matter domination at early times, with no contribution from dark energy. For this point, the secular equation  of the autonomous system (\ref{sysa}) and (\ref{sysb}) is singular. A detailed study of the equations near this critical point shows that the system diverges exponentialy from this critical point, and thus it is an unstable point.  

For the second critical point, $B$, we get $(x=1,y=0)$ and again there isn't any interaction, $\tilde{u}=0$. This corresponds to the pure dark energy dominated era which can be accelerating if $\omega<-1/3$. Therefore depending on the initial conditions, approaching this point, the final state of the universe can be a pure dark energy dominated one in this model. 

Finally the third point, $C$, corresponds to the dark energy-cold dark matter dominated universe. It is only of type I, so that the sign of $\omega$ and $\dot{\varphi}$ is different. It can represent the late time universe as the previous one. For the latter points, $B$ and $C$, one of the eigenvalues is always negative. The second eigenvalue for model $B$ Type II is negative, while for models $B$ Type I and $C$ an appropriate range for $\beta$ can be chosen so that the eigenvalue becomes negative. Therefore model $B$ Type II is stable, while models $B$ Type I and $C$ are stable or saddle depending upon the value of the coupling constant, $\beta$. On the other hand the critical points should be such that $0\le x\le 1$ and $0\le y\le 1$ which for the point $C$ is satisfied provided that $\beta<\sqrt{\frac{3}{1+\omega}}|\omega|$. Therefore the point $C$ is a saddle point. We see that there are only two critical points that admit accelerated solutions, $B$ and $C$. These are in agreement with the solutions  $a$ and $b_M$ of Amendola in \cite{amendo}. The point $C$ is the only accelerated solution in which the ratio of matter and dark energy densities is constant and therefore is a scaling solution. The existence of this critical point can provide the same order of energy densities for both interacting components independently of the initial conditions. This point is also in agreement with the solution $D$ of Bohemer et.al in \cite{bohmer} (See table I of \cite{bohmer}). Restricting ourselves to the positive value of $Q$, which means $\dot\varphi>0$, corresponds to the transfer of energy from quintessence to the dark matter, and this is compatible with the second law of thermodynamics \cite{thermo}. In this case, allowing the observed acceleration expansion of the universe, we have to select the type $I$ solution from any critical point. 
\section{Dark energy--Dark matter ratio}
In what follows, we shall discuss the evolution of the ratio of dark energy to dark matter densities, $r$. To do this, from equations (\ref{sysa}) and (\ref{sysb}),  we obtain \cite{gab}:
\begin{equation}
  \frac{r^\prime}{r} = 3 - \frac{(r+1)}{ry} \tilde{u} \,
  , \label{DEtoDM}
\end{equation}
With the interaction term given by (\ref{z}), the above equation becomes:
\begin{equation}
  \frac{r^\prime}{r} = 3 \mp\tilde{\alpha}\sqrt{x}(1+\frac{1}{r}) \,
  , \label{DEtoDMQ}
\end{equation} 
where $\tilde{\alpha}=\tilde{\beta}\sqrt{3(1+\omega)}$.
To solve this equation, one needs the present value of $r$ as the initial condition. However as the authors of \cite{delcampo} point out it depends on the present time value of Hubble parameter and on the other hand there aren't any model independent observational data for $r(H)$ and $r(z)$. Thus these authors propose to determine the dependence of the Hubble parameter in terms of redshift, $z$, to compare with observational data which are model independent \cite{delcampo}. To do this, using relation (\ref{friedmann1}), equation (\ref{DEtoDMQ}) becomes:
 \begin{equation}
  \frac{dr}{dH} =\frac{2\omega(3r\mp\tilde{\alpha}\sqrt{x}(r+1))}{3H(1+\omega x)}
  , \label{DEDMH}
\end{equation} 
Now combining equation (\ref{DEtoDMQ}) with the expression $H=-dz/[(1+z)dt]$, yields:
 \begin{equation}
  \frac{dr}{dz} =\frac{\omega(3r\mp\tilde{\alpha}\sqrt{x}(r+1))}{(1+z)}
  , \label{DEDMZ}
\end{equation}
Dividing equation (\ref{DEDMZ}) by (\ref{DEDMH}), gives:
\begin{equation}
  \frac{dH}{dz} =\frac{3H(1+\omega x)}{2(1+z)}
  , \label{DHDZ}
\end{equation}
This relation shows that $H(z)$ depends on the interaction term, $Q$, through the dark energy density in this model, see relation (\ref{sysa}). 

Although we have not an exact solution of equations (\ref{DEDMZ}) and (\ref{DHDZ}) in general, the behaviour of $r$ can be studied for extreme limits of the early and the late times. We shall show that provided that $\omega$ and $\beta$ are chosen appropriately, we have solutions that predict the correct behaviour of dark-energy--dark-matter ratio.

 At early times (matter dominated era), for which $x\rightarrow 0$, equation (\ref{DHDZ}) solves to $H\sim (1+z)^{3/2}$, and equation (\ref{DEDMZ}) reads as:
\begin{equation}
\frac{dr}{dz}=\frac{3\omega r}{1+z}
\end{equation} 
with the solution $r\sim (1+z)^{3\omega}$.

On the other hand, at late time, the baryonic matter density is ignorable and thus $x+y\rightarrow 1$, equation (\ref{DEDMZ}) reads as:
\begin{equation}
\frac{dr}{dz}=\frac{3\omega r}{1+z}\mp\frac{\tilde{\alpha}\omega}{1+z}\sqrt{r(1+r)}
\label{s1}
\end{equation} 
which has the following exact solution:
\begin{equation}
1+z=\left (\frac{F_1(r)F_2(r)}{F_1(r_0)F_2(r_0)}\right )^{-\frac{1}{\left(-9+\tilde{\alpha} ^2\right) \omega }} 
\label{s2}
\end{equation}
where:
\begin{equation}
F_1(r)=(1+2 r+2 \sqrt{r (1+r)})^{\pm\tilde{\alpha}}
\label{s3}
\end{equation}
and
\begin{equation}
F_2(r)=\left(\tilde{\alpha}  \left(-6 \sqrt{r (1+r)}+\tilde{\alpha} \right)+r \left(9+\tilde{\alpha} ^2\right)\right)^3
\label{s4}
\end{equation}
and $r_0$ is the current value of $r$.

Now we shall set some assumptions on the asymptotic value and slope of dark energy--dark matter ratio. The late time value $r_\infty=R$ should be such that:
\begin{equation}
R> r_0
\label{s5}
\end{equation}
where $r_0$ is the present value of $r$. Moreover to have a nearly constant dark energy--dark matter ratio at late times we have the following restriction:
\begin{equation}
\left . \frac{dr}{dz}\right|_{r=R, z=Z} << 1
\label{s6}
\end{equation}
where $Z$ tends to $-1$.

Using equation (\ref{s1}), the conditions (\ref{s5}) and (\ref{s6}) lead to:
\begin{equation}
\tilde{\alpha}<3\sqrt{\frac{R}{1+R}}
\end{equation}
and 
\begin{equation}
\omega << \frac{1+Z}{3R-\tilde{\alpha}\sqrt{R(1+R)}}
\end{equation}

\begin{figure}[htb]
\begin{center}
\epsffile{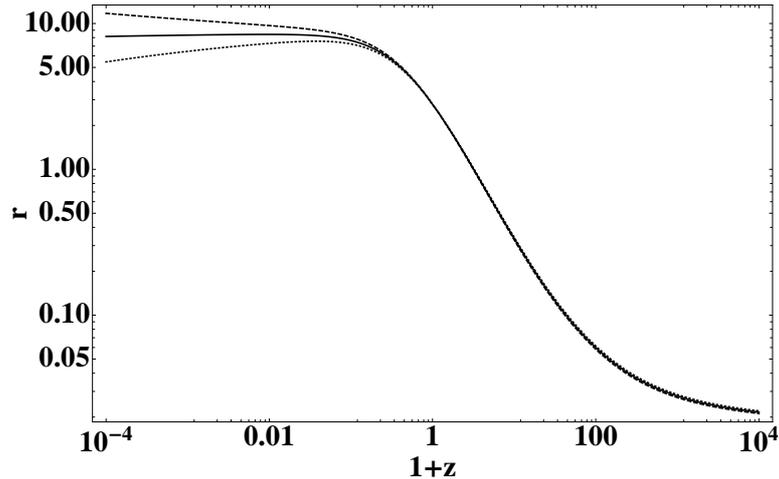}
\end{center}
\caption{Dark energy--Dark matter ratio as a function of redshift. $\omega$ is chosen to be $-0.5$ and $\beta$ equals to $1.14$, $1.19$ and $1.16$ for dashed, dotted and thick lines respectively. $r$ at present time is chosen to be $7/2.5$.}
\label{f1}
\end{figure}

\begin{figure}[htb]
\begin{center}
\epsffile{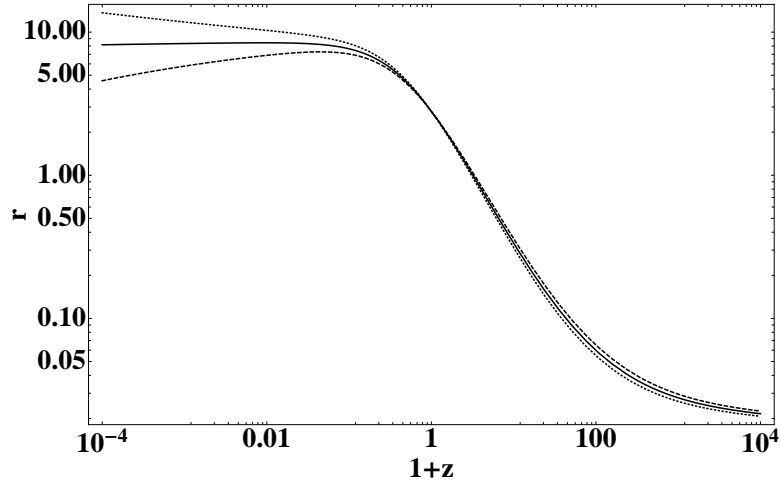}
\end{center}
\caption{Dark energy--Dark matter ratio as a function of redshift. $\beta$ is chosen to be $1.16$ and $\omega$ equals to $-0.49$, $-0.51$ and $-0.5$ for dashed, dotted and thick lines respectively. $r$ at present time is chosen to be $7/2.5$.}
\label{f2}
\end{figure}

The evolution of dark energy--dark matter ratio with respect to the redshift is shown in figures (\ref{f1}) and (\ref{f2}). In figure (\ref{f1}) graphs for different values of $\beta$ and a fixed $\omega$ are drawn while in figure (\ref{f2}) graphs for different values of $\omega$ and a fixed $\beta$ are considered. We see that this model leads to a constant unstable $r$ at late times and to a lower constant stable $r$ for early times. This shows a similar behaviour to what is obtained in \cite{gab,delcampo} for two broad classes of interacting models of dark energy.

\begin{figure}[htb]
\begin{center}
\epsffile{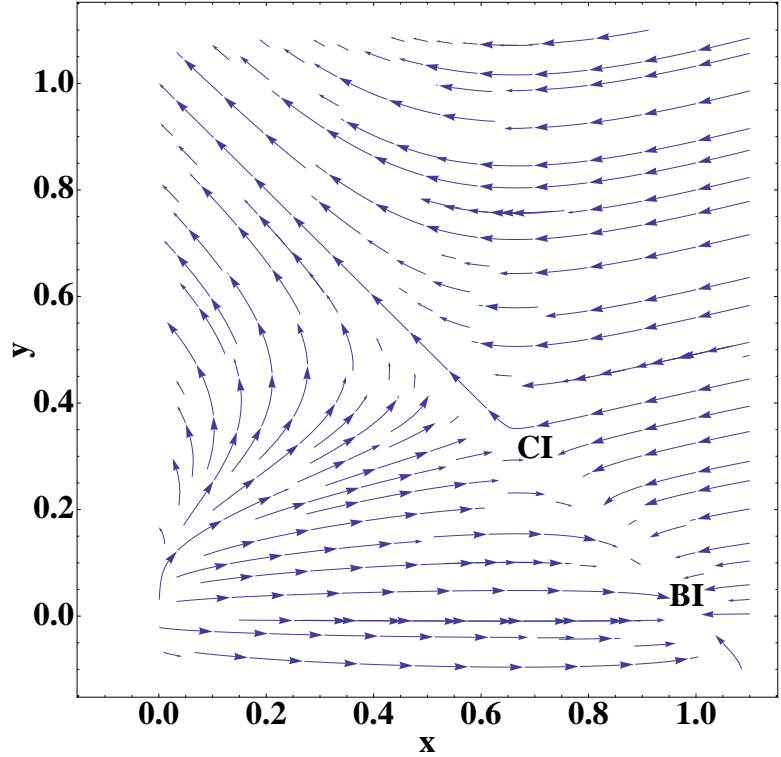}
\end{center}
\caption{Phase diagram of the type $I$ system for $\omega=-0.5$ and $\tilde{\beta}=2$. $BI$ and $CI$ denote points $B$ and $C$ of type $I$.}
\label{f3}
\end{figure}

\begin{figure}[htb]
\begin{center}
\epsffile{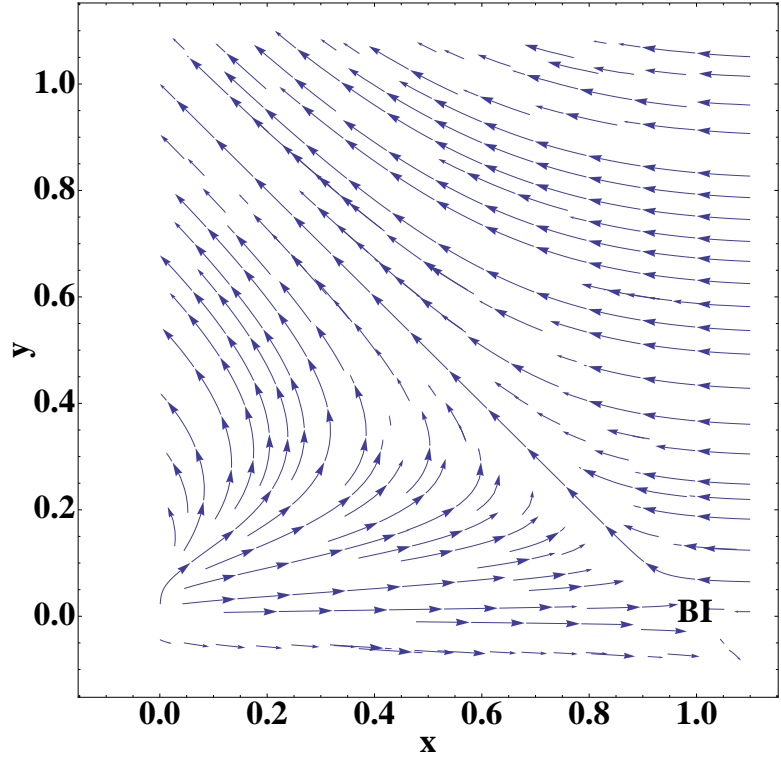}
\end{center}
\caption{Phase diagram of the type $I$ system for $\omega=-0.5$ and $\tilde{\beta}=2.7$. $BI$  denotes point $B$ of type $I$. The point $CI$ is now out of acceptable range of $x$ and $y$.}
\label{f4}
\end{figure}

\begin{figure}[htb]
\begin{center}
\epsffile{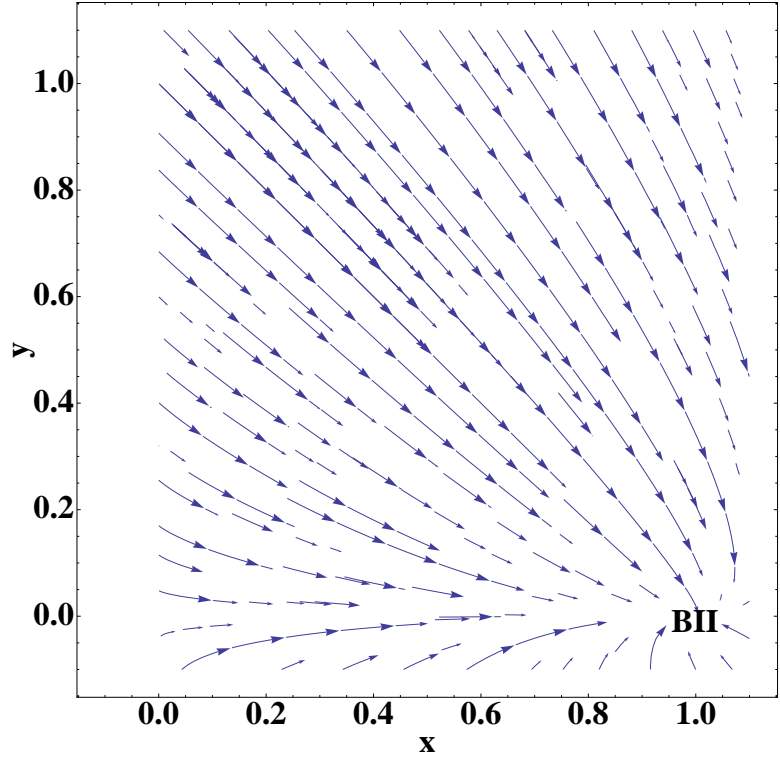}
\end{center}
\caption{Phase diagram of the type $II$ system for $\omega=-0.5$ and $\tilde{\beta}=2$. $BII$  denotes point $B$  of type $II$.}
\label{f5}
\end{figure}

In figures (\ref{f3}), (\ref{f4}) we have plotted the phase space diagrams of type $I$ model (equations (\ref{sysa}), (\ref{sysb}) and (\ref{z})). Figure (\ref{f5}) is the corresponding plot for type $II$ model. 

In figure (\ref{f3}) the stability condition of $BI$ point is satisfied and thus the point $CI$ is a saddle point as it is expected from table 1. Changing the value of $\tilde{\beta}$ such that the stability condition of $BI$ point is not satisfied leads to figure (\ref{f4}) and one can see that the critical point $CI$ is out of acceptable range of $x$ and $y$.

Finally as it can be seen in figure (\ref{f5}), for a type $II$ solution the point $BII$ is an attractor point.

\epsfxsize=4in
\epsfysize=4in
\epsfxsize=4in
\epsfysize=4in
\epsfxsize=4in
\epsfysize=4in
\epsfxsize=4in
\epsfysize=4in
\epsfxsize=4in
\epsfysize=4in

{A question that arises here is that which kind of potential leads to the derived solution. 
In order to reconstruct the potential, one can use the relations $V=(1-\omega)\rho_\varphi/2$, $\dot{\varphi}=\sqrt{(1+\omega)\rho_\varphi}$ and equation (\ref{rhophidot}) to obtain:}
\begin{equation}
\frac{dV}{d\varphi}=-\frac{1-\omega}{2}\left( 3H\sqrt{(1+\omega)\rho_\varphi}+\beta\rho_c\right)
\label{a}
\end{equation}
{$\rho_\varphi$ in the above equation can be substituted by integration equation (\ref{rhocdot}) leading to\cite{epl-khodemoon}:}
\begin{equation}
\rho_c=\rho_{c0}\left(\frac{a}{a_0}\right)^3\exp\left(\beta(\varphi-\varphi_0)\right)
\end{equation}
{where a subscript 0 denotes the current values of quantities.}

{At late times, $z\sim -1$, $\rho_\varphi/\rho_c\sim R$ and using the equations (\ref{rhocdot}) and (\ref{rhophidot}) we get:}
\begin{equation}
\rho_c=\rho_{c0}\left(\frac{a_0}{a}\right)^{3\left( 1+\frac{\omega R}{1+R} \right)}
\end{equation}
{Inserting the above equation in the equation(\ref{friedmann}) leads to:}
\begin{equation}
H^2=\frac{1+R}{3}\rho_{c0}\exp\left(\beta\left(1+\frac{1+R}{\omega R}\right)(\varphi-\varphi_0)\right)
\end{equation}
{and thus equation (\ref{a}) can be integrated to:}
\begin{equation}
V\sim\exp\left(\beta\left(1+\frac{1+R}{\omega R}\right)(\varphi-\varphi_0)\right)
\end{equation}
{For early times, $\rho_\varphi$ is small and thus $\dot\varphi\sim 0$ and $\varphi\sim\varphi_0$ and therefore the above potential can be expanded up to first order. Using this approximate potential one can check that the Friedmann equations and the equations governing densities are satisfied.}

\section{Conclusion}
In this paper we have investigated non--minimal quintessence model of dark energy. Following \cite{gab}, we have defined a new variable, $N=-\omega \ln a$, which reduces the parameter space into a two dimmensional space.  Our results are in agreement with those of \cite{amendo,bohmer}. We found some range of the interaction parameter $\beta$ given by the stability criteria of the critical points, and we presented an exact solution of dark--energy to dark--matter ratio for the early and late times.

According to the early time behaviour ($r\sim (1+z)^{3\omega}$) the model behaves like $\Lambda$CDM model. On the other hand it can provide a finite dark--energy to dark--matter ratio at late times and thus it can alleviate the coincidence problem.
In fact it is shown here that provided the strength of the interaction ($\beta$) is chosen appropriate, this interacting model can predict the expected behaviour for the dark-energy to dark-matter ratio with a constant $\omega$.

\textbf{Acknowledgement} This work is partly supported by a grant from university of Tehran and partly by a grant from center of excellence of department of physics on the structure of matter.


\begin{thebibliography}{99}
\bibitem{Riess}
Riess, A. G. et. al,  \textit{Astron. J.}, \textbf{116}, 1009, 1998;
Perlmutter, S. et al, \textit{Astrophys. J.}, \textbf{517}, 565, 1999.
\bibitem{sami}
Copeland, E. J., Sami,  M., and Tsujikawa, S., \textit{Int . J. Mod. Phys. D}
\textbf{15}, 1753, 2006; Peebles, P. J. E., and Ratra,  B., \textit{Rev. Mod.
Phys.}, \textbf{75}, 559, 2003.
\bibitem{carroll}
Padmanabhan, T. ,\textit{Phys. Rept.}
\textbf{380}, 235, 2003 ;  
Carroll, S. M., Press, W. H. and Turner, E. L. , \textit{Annu. Rev. Astron. Astrophys.}
\textbf{30}, 499, 1992 ;
Carroll, S. M.,  \textit{Liv. Rev. Rel.}
\textbf{4}, 1, 2001.
\bibitem{quintessence}
Ratra B., and Peebles P. J. E, \textit{Phys. Rev D} \textbf{37}, 3406, 1988; 
Peebles P. J. E., and Ratra, B., \textit{Astrophys. J. Lett.} \textbf{325}, L17, 1988;
Caldwell R. R., Dave R. , Steinhardt P. J. , \textit{Phys. Rev. Lett.}
\textbf{80}, 1582, 1998.
\bibitem{coincidence}
Weinberg, S., \textit{Cosmology}, Oxford University Press, New
York, Sec. 1.12, 2008;  Carroll, S. M., \textit{AIP Conf. Proc.} \textbf{743}, 16,  2005; Franca, U.,  Rosenfeld, R. \textit{JHEP} \textbf{10}, 015, 2002. 
\bibitem{amen}
Toccin-Valentini D. and Amendola, L. , \textit{Phys. Rev. D.}
\textbf{65}, 063508, 2002 ;
Zimdahl,W. , Pavon, D. , \textit{Gen. Relativ. Grav.}
\textbf{35}, 413, 2003.
\bibitem{2olivares}
Chimento, L. P., Jakubi, A. S. , Pavon, D., and Zimdahl, W. \textit{Phys. Rev. D.}
\textbf{67}, 083513, 2003; 
Zimdahl, W., Pavon, D., and Chimento, L. P.,  \textit{Phys. Lett. B.}
\textbf{521}, 133, 2001.
\bibitem{campo}
Campo, S. D.,  Herrera, R. , Olivares, G. , Pavon, D. \textit{Phys. Rev. D.}
\textbf{74}, 023501, 2006.
\bibitem{oliv}
Olivares, G. , Atrio-Barandela, F. , Pavon, D. \textit{Phys. Rev. D.}
\textbf{71}, 063523, 2005;
Barrow, J. D., and Clifton, T., \textit{Phys. Rev. D.}
\textbf{73}, 103520, 2006;
Gumjudpai, B., Naskar, T. ,Sami, M., and Tsujikawa, S. \textit{Phys. Rev. D.}
\textbf{73}, 103520, 2006.
\bibitem{gab}
Caldera-Cabral, G. , Maartens, R. , and Ure\~{n}a-L\'{o}pez, L. A., \textit{Phys. Rev. D.}
\textbf{79}, 063518, 2009;
\bibitem{Q}
Quartin, M. , Calvao, M. O., Joras, S. E. , Reis, R. R. R., and Waga, I.\textit{J. Cosmol. Astropart. Phys.}
\textbf{05}, 007, 2008.
\bibitem{amend}
Amendola, L., Quartin, M., Tsujikawa, S., and Waga, I. \textit{Phys. Rev. D.}
\textbf{74}, 023525, 2006.
\bibitem{guo}
Guo, Z. K., Ohta, N., and, Tsujikawa, S., \textit{Phys. Rev. D.}
\textbf{76}, 023508, 2007;
Zimdahl,W. , Pavon, D. and Chimento, L. P. ,  \textit{Phys. Lett. B}
\textbf{521}, 133, 2001;
Quercellini, C., Bruni, M., Balbi, A., and Pietrobon, D., \textit{Phys. Rev. D.}
\textbf{78}, 063527, 2008
\bibitem{epl-khodemoon}
A. Shojai, and F. Shojai, \textit{Euro. Phys. Lett.}, \textbf{88}, 30002, 2009.
\bibitem{amendo}
Amendola, L. \textit{Phys. Rev. D.}
\textbf{62}, 043511, 2000;
Amendola, L. \textit{Phys. Rev. Lett.}
\textbf{86}, 196, 2001. 
\bibitem{bohmer}
Bohmer, C. G., Caldera-Cabral, G. , Lazkoz, R., and Maartens, R.  \textit{Phys. Rev. D.}
\textbf{78}, 023505, 2008.
\bibitem{s1}
Roshan M., Shojai F., \textit{Phys. Rev. D.}
\textbf{79}, 103510, 2009;
Roshan M., Shojai F., \textit{Phys. Rev. D.}
\textbf{80}, 043508, 2009.
\bibitem{3bohmer}
Wetterich C., \textit{Astron. Astrophys.} \textbf{301}, 321, 1995;
 Holden, D. J., and Wands, D., \textit{Phys. Rev. D.}
\textbf{61}, 043506, 2000; 
\bibitem{6bohmer}
 Huey, G., and Wandelt, B. D., \textit{Phys. Rev. D.}
\textbf{74}, 023519, 2006;
Das,S., Corasaniti, P. S., and Khoury, J., \textit{Phys. Rev. D.}
\textbf{73}, 083509, 2006; 
Bean, R. and Magueijo, J. \textit{Phys. Lett. B}
\textbf{517}, 177, 2001;
\bibitem{energy}
Amendola, L. and Tsujikawa,  \textit{Dark Energy: Theory and Observation}, Cambridge University Press, New
York, 2010;
\bibitem{11boom}
Damour, T., Gibbons, G. W. and Gundlach, C., \textit{Phys. Rev. Lett.}
\textbf{64}, 123, 1990;
\bibitem{tor}
Torres, D. F., \textit{Phys. Rev. D.}
\textbf{66}, 043522, 2002;
\bibitem{mata}
Matarrese, S., Baccigalupi, C. and Perrotta, F., \textit{Phys. Rev. D.}
\textbf{70}, 061301, 2004; 
\bibitem{sahni}
Sahni, V. and habib, S., \textit{Phys. Rev. Lett.}
\textbf{81}, 1766, 1998;
\bibitem{chom}
Sahni, V., and Starobinsky, A., \textit{Int. J. Mod. Phys. D}, \textbf{9}, 373, 2000.
\bibitem{thermo}
Pavon, D., and Wang, B.  \textit{Gen. Relativ. Gravit.}
\textbf{41}, 1, 2009;
\bibitem{delcampo}
Campo, S. D.,  Herrera, R. , Pavon, D. \textit{Phys. Rev. D.}
\textbf{78}, 021302, 2008.
\end{thebibliography}
\end{document}